\documentclass[iop,apj]{emulateapj}
\usepackage{amsmath,amssymb, amsthm,amstext} 
\usepackage{natbib}
\usepackage{graphicx}
\usepackage{color}
\usepackage{array}
\usepackage{bm}
\usepackage[breaklinks,colorlinks,citecolor=blue]{hyperref}

\newtheorem{claim}{Claim}

\def\nn{\nonumber}
\def\st{\sin\theta}
\def\ct{\cos\theta}
\def\sst{\sin^2\theta}

\def\Ar{A_{\phi,r}}
\def\Arr{A_{\phi,rr}}
\def\Am{A_{\phi,\mu}}
\def\Amm{A_{\phi,\mu\mu}}
\def\Ah{A_{\phi,\theta}}
\def\be{\begin{equation}}
\def\ee{\end{equation}}
\def\ben{\begin{eqnarray}}
\def\een{\end{eqnarray}}

\def\A0{A_{\phi}^{(0)}}
\def\e{{\bf e}}

\begin{document}
\title{Analytic properties of force-free jets in the Kerr spacetime -- II}

\slugcomment{Submitted to \apj}

\author{Zhen Pan}
\affil{Department of Physics, University of California, One Shields Avenue, Davis, CA 95616, USA}
\email{zhpan@ucdavis.edu}

\author{Cong Yu}
\affil{Yunnan Observatories, Chinese Academy of Sciences, Kunming 650011, China;\\
Key Laboratory for the Structure and Evolution of Celestial Objects, Chinese Academy of Sciences, Kunming 650011, China}
\email{cyu@ynao.ac.cn}

\shorttitle{Analytic properties of force-free jets in the Kerr spacetime }
\shortauthors{Z. Pan and C. Yu}

\begin{abstract}
We reinvestigate the structure of a steady axisymmetic force-free magnetosphere around a Kerr black hole (BH).  The BH magnetosphere structure is governed by a second-order differential equation of $A_\phi$ depending on two `free' functions $\Omega$ and $I$, where $A_\phi$ is the $\phi$ component of the vector potential of the electromagnetic field, $\Omega$ is the angular velocity of the magnetic field lines and $I$ is the poloidal electric current. While the two functions $\Omega$ and $I$ are not arbitrarily given, which need to be self-consistently determined along with the differential equation. Based on the perturbation approach we proposed in paper I \citep{Pan2015a}, in this paper,  we self-consistently sort out two boundary conditions governing $\Omega$ and $I$, and interpret these conditions mathematically and physically. Making use of the boundary conditions, we prove that all magnetic field lines crossing the infinite-redshift surface also penetrate the event horizon. Furthermore, we argue that the BH Meissner effect does not work in force-free magnetosphere due to the perfect conductivity. 
\end{abstract}


\bigskip\bigskip

\section{Introduction}
The Blandford-Znajek (BZ) mechanism \citep{Blandford1977d} is believed to be one of most efficient way to extract rotation energy from a spinning black hole (BH), which operates in BH systems on all mass scales, from the stellar-mass BHs of gamma ray bursts to the supermassive BHs of active galactic nuclei. In the past decade, we have gained more and more understanding of the BZ mechanism from  general relativistic magnetohydrodynamic (GRMHD) simulations  \citep{Komissarov2001,Komissarov2004e, Komissarov2004,Komissarov2005, Semenov2004, McKinney2004f, McKinney2005, Komissarov2007d, 2008MNRAS.388..551T, Tchekhovskoy2010,  2011MNRAS.418L..79T, Palenzuela2011, 2012MNRAS.423L..55T, Penna2013a, 2013Sci...339...49M}, analytic studies \citep{Tanabe2008, Beskin2013, Pan2014, Pan2015a, Pan2015b, Gralla2015, Yang2015, Penna2015} and numerical solutions \citep{Palenzuela2010, Contopoulos2013, Nathanail2014}. Various studies converge to a common picture of how the BZ mechanism works: poloidal magnetic field $\bm{B_{\rm p}}$, threads  non-rotating BHs, while spinning BHs distorts those field lines and induces poloidal electric fields $\bm{E_{\rm P}}$ and toroidal magnetic fields $\bm{B_{\rm T}}$, thus an outward Poynting flux $\bm{E_{\rm P} \times B_{\rm T} }$ is generated along the magnetic field lines threading the spinning BH, and the rotation energy of the spinning BHs is extracted in the form of Poynting flux. 

While there were some clues that BH rotation tends to expel magnetic field lines out of the event horizon, which is known as the BH Meissner effect. \citet{Wald1974} gave the vacuum solution for the electromagnetic field when a Kerr BH is placed in an originally uniform magnetic field aligned along the BH rotation axis. For this solution, \citet{King1975} found all magnetic field expelled out of the horizon of an extremal Kerr BH.  \citet{Bicak1976,Bicak1985} generalized this result and showed that for all steady axisymmetric vacuum solutions, no non-monopole component of magnetic flux penetrates the event horizon if the BH is extremal. Consequently, there were some concerns that the BZ mechanism would be quenched if the BH Meissner effect works in real astrophysical environment. But such effect was never seen in previous GRMHD simulations. Different field line types in BH magnetosphere found in simulations were discussed in \citet{Blandford2002, Hirose2004, Mckinney2005a}.  

To explain the absence of the Meissner effect in time-dependent simulations, quite a few possibilities have been discussed. Non-axisymmetry was proposed \citep{Bicak1985, Penna2014} to evade the effect, but it does not explain the absence of the Meissner effect in axisymmetic simulations. \citet{Komissarov2007d} argued that it is conductivity that breaks the BH Meissner effect.  As they pointed out, the Wald solution and all vacuum solutions constructed by \citet{Bicak1985} are electric current fixed. Introducing conductivity completely changes the current distribution and the magnetic field configurations.  Recently, \citet{Penna2014} put forward that the BH Meissner effect is a geometry effect, which allows purely (split) monopole magnetic fields at the horizon of an extremal Kerr BH and all other field components would be expelled. But the conclusion seems controversial: taking the most well studied asymptotically split-monopole solution as an example \citep{Blandford1977d, Tanabe2008, Pan2015a, Pan2015b}, the magnetic field is purely monopole only in the Schwarzschild spacetime.  BH rotation distorts the field configuration and the distortion becomes larger for faster spinning BHs. So the magnetic fields are no longer purely monopole in the Kerr spacetime, especially for extremal Kerr BHs. \citet{Takamori2011} also argued that higher multipole components may be superposed on the monopole component on the event horizon even for extremal Kerr BHs, if an electric current exists.

In this paper, we reinvestigated the structure of a steady axisymmetic force-free magnetosphere in the Ker spacetime and the BH Meissner effect. It is known that the magnetosphere structure is governed by the general relativistic Grad-Shafranov (GS) equation. A perturbation approach was proprosed to self-consistently solve $A_\phi$ and determine $I$ and $\Omega$ \citep{Pan2015a}. Based on the perturbation approach, we find that all known solutions to the GS equation satisfy simple boundary/constraint conditions. With these conditions, we prove that all magnetic field lines crossing the infinite redshift surface also penetrate the event horizon. Hence, if the BH Meissner effect operates in a force-free magnetosphere, all field lines would be expelled out of the event horizon and out of the infinite-redshift surface. We argue that this will not happen.

The paper is organized as follows. Basic equations governing a steady axisymmetric force-free magnetosphere in the Kerr spacetime are summarized in Section \ref{sec:basic}. In Section \ref{sec:boundary}, we first introduce the analytic approach to solve the GS equation and specify boundary conditions governing $I$ and $\Omega$, then interpret these boundary conditions mathematically and physically. Based on the boundary conditions, we study the structure of magnetic field lines near the central BH in Section \ref{sec:field lines}. In Section \ref{sec:meissner}, we argue that the BH Meissner effect does not work in a force-free magnetosphere. Conclusions are presented in Section \ref{sec:conclusion}.

\section{Basic equations}
\label{sec:basic}

In this section, we sketch the basic equations governing steady axisymmetric force-free electromagnetic field around Kerr black holes
(see \citet{Pan2014} and references therein for more details). We adopt the Kerr-Schild (KS) coordinate \citep{1963PhRvL..11..237K,McKinney2004f}
with the line element
\ben
ds^2
&& = -\left( 1-\frac{2r}{\Sigma} \right)dt^2 + \left( \frac{4
r}{\Sigma} \right) dr dt + \left(1+\frac{2r}{\Sigma} \right) dr^2\nn\\
&& 
+\Sigma d\theta^2 - \frac{4 a r \sin^2\theta}{\Sigma} d\phi dt
- 2 a \left(1+\frac{2r}{\Sigma}\right) \sin^2\theta d\phi dr  \nn\\
&&
+\frac{\beta}{\Sigma}\sst d\phi^2 \ ,
\een
where $\Sigma=r^2+a^2\cos^2\theta$, $\Delta=r^2-2r+a^2$, $\beta = (r^2+a^2)^2-a^2\Delta\sst$ and the square root of determinant $\sqrt{-g}=\Sigma\sin\theta$.
Steady axisymmetric force-free electromagnetic fields in the Kerr spacetime are determined by three functions $A_\phi$, $\Omega(A_\phi)$, $I(A_\phi)$, 
where $A_\phi$ is the $\phi$ component of electricmagnetic potential, $\Omega(A_\phi)$ is angular velocity of magnetic field lines and $I(A_\phi)$ is poloidal electric current flowing into the central BH. Non-trivial components of Faraday tensor $F_{\mu\nu}\equiv \partial_\mu A_\nu-\partial_\nu A_\mu$ in the Kerr-Schild coordinate could be expressed as follows \citep{Blandford1977d,McKinney2004f} 
\be 
F_{r\phi} = -F_{\phi r}=A_{\phi,r} \ , 
F_{\theta\phi} = - F_{\phi\theta} = A_{\phi,\theta} \ , 
\ee 
\be F_{tr} = -F_{rt} = \Omega A_{\phi,r} \ , 
F_{t\theta}= - F_{\theta t} = \Omega A_{\phi,\theta} \ , 
\ee 
\be
F_{r\theta} = -F_{\theta r} = \sqrt{-g}B^{\phi} \ , 
\ee 
where 
\be
B^\phi = - \frac{I \Sigma + (2 \Omega r - a) \sin\theta
A_{\phi,\theta}} {\Delta \Sigma \sin^2\theta}\ .
\ee 
The energy conservation equation of the electromagnetic field reads 
\ben
\label{eq:GS} 
&-&\Omega \left[(\sqrt{-g}F^{tr})_{,r} +
(\sqrt{-g}F^{t\theta})_{,\theta} \right] + F_{r\theta}I'(A_\phi) \nn\\
&&+ \left[(\sqrt{-g}F^{\phi r})_{,r} +
(\sqrt{-g}F^{\phi\theta})_{,\theta} \right] = 0 \ , \een 
which is also known as the GS equation.

In the Schwarzschild spacetime, $I = \Omega = a = 0$, the GS equation above is
simplified as 
\be 
\mathcal L A_\phi =0, 
\ee 
where the operator 
\be
\mathcal L\equiv\frac{1}{\sin\theta}\frac{\partial}{\partial
r}\left(1-\frac{2}{r}\right)\frac{\partial}{\partial r}
 +\frac{1}{r^2}\frac{\partial}{\partial \theta}\frac{1}{\sin \theta}\frac{\partial}{\partial \theta} ,
\ee
and its Green's function $G(r,\theta; r_0, \theta_0)$ defined by 
$ \mathcal LG(r,\theta; r_0,\theta_0) = \delta(r-r_0) \delta(\theta-\theta_0), $ is available \citep{1974PhRvD..10.3166P,Blandford1977d}.

In the Kerr spacetime, the GS equation explicitly expresses as 
\ben
\label{eq:GSe}
&&
\left[\left(\frac{\beta}{\Sigma}\right)\Omega^2 
-\frac{4ra}{\Sigma}\Omega
- \frac{1}{\sst}\left(1-\frac{2r}{\Sigma}\right)\right] \Arr \nn\\
&&
+ \Bigg[ \left(\frac{\beta}{\Sigma}\right)_{,r}\Omega^2 
-\left(\frac{4ra}{\Sigma}\right)_{,r} \Omega
+\frac{1}{\sst}\left(\frac{2r}{\Sigma}\right)_{,r}\Bigg]\Ar\nn\\
&&
+ \left[ \left(\frac{\beta}{\Sigma}\right) \Omega-\frac{2ar}{\Sigma}\right]\Omega'\Ar^2 \nn\\
&&
+\left[\left(1+\frac{2r}{\Sigma}\right)\Omega^2(\sst\Am)_{,\mu}-\frac{1}{\Sigma}\Amm \right]\nn\\
&&
+\left(1+\frac{2r}{\Sigma}\right)\Omega\Omega'\sst\Am^2 \nn\\
&&
+(2r\Omega^2\sst -1)\left(\frac{1}{\Sigma}\right)_{,\mu}\Am \nn\\
&&
+\frac{\Sigma}{\Delta\sst} 
\left(\frac{(2r\Omega-a)\sst}{\Sigma}\right) \left(\frac{(2r\Omega-a)\sst \Am}{\Sigma}\right)_{,\mu}\nn\\
&&
- \frac{\Sigma }{\Delta\sst} II' = 0,
\een
where $\mu\equiv\ct$ and  the prime designates the derivative with respect to $A_\phi$. \citet{Contopoulos2013} and \citet{Nathanail2014} also obtained the GS equation with slightly different notations. Generally speaking, $\Omega$ and $I$ need to be specified by two boundary/constraint conditions, which are not arbitrary and  need to be solved self-consistently together with the GS equation.

\section{Boundary conditions}
\label{sec:boundary}
In this section, we start with introducing the perturbation approach of solving the GS equation and determining the boundary conditions self-consistently.  Then we interpret these boundary conditions mathematically and physically.

\subsection{Perturbation method}
The Schwarzschild metric solution to the GS equation (\ref{eq:GS}) writes as
\be\label{eq:schzsol}
\Omega_0 = 0, \quad I_0 = 0, \quad \mathcal LA_0=0.
\ee
For the corresponding Kerr metric solution, we define $\omega = \Omega(A_\phi)|_{r\rightarrow\infty}$, $i =  I(A_\phi)|_{r\rightarrow\infty}$, and expand them in series,
\ben
A_\phi &=& A_0 + a^2 A_2 + a^4 A_4 +  ..., \nn\\
\omega &=& a \omega_1 + a^3\omega_3 + a^5\omega_5 + ...,\nn\\
i &=& a i_1 + a^3 i_3 + a^5i_5+ ... .
\een
As in \citet{Pan2015a}, $\Omega$ and $I$ could be expressed in terms of $\omega_{1,3,5, ...}$ and $i_{1,3,5, ...}$ respectively.
With the above notations, the GS equation (\ref{eq:GS}) could be decomposed as a set of linear equations
\be 
\mathcal L A_n(r,\theta) = S_n(r,\theta; i_{n-1},\omega_{n-1}) ,
\ee
where $n=2, 4, 6, ...$ . For each equation, two functions $i$ and $\omega$ are determined by the horizon regularity condition and the convergence constraint \citep{Blandford1977d, Pan2014}, where the horizon regularity condition requires $B^\phi$ to be finite on the horizon $r=r_+$ and writes as 
\be
\label{eq:znajek}
I =- \frac{(2r\Omega-a)\st \Ah}{\Sigma} , 
\ee
and the convergence constraint requires that all solutions $A_n$ should be convergent from horizon to infinity. With $i_{n-1}$ and $\omega_{n-1}$ determined by these two conditions, $A_n$ is obtained by the integral  
\be 
A_n(r,\theta) = \int_2^\infty dr_0 \int_0^\pi d\theta_0 S_n(r_0,\theta_0) G(r,\theta; r_0,\theta_0). 
\ee

For the asymptotically monopole solution, $A_0=-\ct$, we found $i_n = \omega_n \sst$ for any integer $n$, hence $I = \Omega\sst|_{r\rightarrow\infty}$. Note that both $I$ and $\Omega$ are functions of $A_\phi$, so $I/\Omega$ is also a function of $A_{\phi}$. We denote this function as $ I/\Omega \equiv f(A_\phi)$, which can be readily determined at infinity. It is known that $I / \Omega |_{r\rightarrow\infty} = \sin^2\theta$ and $A_\phi|_{r\rightarrow\infty} = -\cos\theta$, and so we have $f(-\cos\theta) = \sin^2\theta$ or $f(x) = 1-x^2$. 
Finally, we arrive at the \emph{exact} constraint relation \citep{Pan2015a}
\be
\label{eq:c1}
I = \Omega(1-A_\phi^2).
\ee

Applying the above perturbation approach to the asymptotically uniform vertical magnetic field $A_0 = r^2\sst$, we find the \emph{exact} constraint
\be
\label{eq:c2}
\quad I = 2\Omega A_\phi.
\ee
Applying it to the paraboloidal magnetic field $A_0 = r(1-\mu) + 2(1+\mu)(1-\ln(1+\mu))-4(1-\ln2)$ \citep{Blandford1977d}, we also find 
\be
\label{eq:c3}
I = 2\Omega A_\phi.
\ee

To summarize, the three known solutions satisfy the constraints in the form of
\be
\label{eq:IOmega}
I = \Omega\times \mathcal{F}(A_\phi),
\ee
where $\mathcal{F}(A_\phi)$ depends on the configuration of magnetic field and
is of $\sim O(A_\phi)$.

\subsection{Mathematical interpretation}

In the limit of $r\rightarrow\infty$, we expect $A_\phi\rightarrow A_0, \Omega\sim \Omega' \sim I \sim I '\sim O(1)$, and $\Ar\sim O(A_\phi/r),\Arr\sim O(A_\phi/r^2)$. Thus the GS equation (\ref{eq:GSe}) consists of terms approaching to zero at infinity and terms not. The convergence constraint requires the summation of terms not approaching to zero vanishes, so collecting all these terms we have 
\be
\label{eq:infty}
\Omega(\Omega r^2 \Ar)_{,r} + \Omega(\Omega \sst \Am)_{,\mu}=\frac{1}{\sst}II' \Big|_{r\rightarrow\infty}\ ,
\ee
where we have used the unperturbed GS equation in the Schwarzschild spacetime $\mathcal L A_0 =0$. 
The form of the above equation implies that Eq.(\ref{eq:IOmega}) is generally true, i.e. $I\propto\Omega$ despite field configuration.\\

For asymptotically monopole magnetic fields $A_\phi|_{r\rightarrow\infty}= A_0 = -\ct$, more accurately $A_\phi = A_0 + O(A_0/r)$ at large distance, therefore Eq.(\ref{eq:infty}) reads as 
\be
\Omega\sst(\Omega \sst )' = II'\Big|_{r\rightarrow\infty}\ .
\ee
It is easy to get $I= + \Omega \sst|_{r\rightarrow\infty}$, where the positive sign is assigned to enable Poynting flux flow out at infinity (see next subsection for details). Consequently we reproduce Eq.(\ref{eq:c1}), $I=\Omega(1-A_\phi^2)$.

For asymptotically uniform vertical magnetic fields $A_\phi|_{r\rightarrow\infty}= A_0 = r^2\sst$, Eq.(\ref{eq:infty}) reads as  
\be
4\Omega A_\phi (\Omega' A_\phi +\Omega) = II'\Big|_{r\rightarrow\infty}.
\ee
Hence, we reproduce Eq.(\ref{eq:c2}), $I = 2\Omega A_\phi$. 

For paraboloidal magnetic fields, we could also reproduce Eq.(\ref{eq:c3}), $I = 2\Omega A_\phi$, from Eq.(\ref{eq:infty}). \\

In a similar way, we require all terms in Eq.(\ref{eq:GSe}) to be finite in the limit of $r\rightarrow r_+$ . The requirement yields
\be
\frac{(2r\Omega-a)\sst}{\Sigma} \left(\frac{(2r\Omega-a)\sst \Am}{\Sigma}\right)_{,\mu}-II'=0.
\ee
Consequently, we have
\be
I = \frac{(2r\Omega-a)\sst \Am}{\Sigma} , \nn
\ee
which is exactly the Znajek regularity condition (\ref{eq:znajek}). The positive sign is chosen here to enable the Poynting flux measured by physical observers at horizon flow inward (see next subsection for details).

\subsection{Physical interpretation}
From mathematical viewpoint, the boundary conditions are results of convergence requirement of solution $A_\phi$. While in the viewpoint of physics, the boundary conditions are in fact radiation conditions at infinity. \citet{Nathanail2014} proposed that the event horizon is an ``inner infinity'' similar to the outer infinity and they showed that the Znajek regularity condition (\ref{eq:znajek}) is equivalent to the ingoing radiation condition at inner infinity. \citet{Penna2015} generalized the Znajek regularity condition to outer infinity and showed that it is equivalent to the outgoing radiation condition. Here we apply the radiation conditions on different magnetic field configurations and show that they are equivalent to the constraint relations (\ref{eq:c1}-\ref{eq:c3}). 

For comparison with previous works, we turn to use Boyer-Lindquist (BL) coordinate \footnote{The  KS coordinate and the BL coordinate are connected by \citep{McKinney2004f} $\frac{\partial r\phantom{\mu} [{\rm KS}]}{\partial x^\mu [{\rm BL}]}=\delta^r_{\ \mu}$ and $\frac{\partial \theta\phantom{\mu} [{\rm KS}]}{\partial x^\mu [{\rm BL}]}=\delta^\theta_{\ \mu}$, therefore the functions $I(r,\theta), \Omega(r,\theta), A_\phi(r,\theta)$ defined in the KS coordinate do not change their forms when expressed in the BL coordinate. Consequently, the boundary/constraint conditions [Eqs.(\ref{eq:znajek},\ref{eq:IOmega})] involving $r,\theta$ coordinate only do not change their forms either.}
\ben
\label{eq:boyer}
ds^2 
&=& 
-\left(1-\frac{2r}{\Sigma}\right) dt^2 -\frac{4ar\sst}{\Sigma} dt d\phi+\frac{\Sigma}{\Delta} dr^2 \nn\\
&&
+ \Sigma d\theta^2 + \frac{\beta}{\Sigma}\sst d\phi^2
\een
in the following discussion. For physical understanding, we introduce the tetrad carried by zero-angular momentum observers (ZAMOs) in the Kerr space time. In the Boyer-Lindquist coordinate, the tetrad is written as \citep{Bardeen1972}
\ben
\label{eq:tetrad}
(\e_{\hat t})^\mu &=& \sqrt{\frac{\beta}{\Sigma\Delta}}(1, 0, 0, \frac{2ar}{\beta}), \nn\\
(\e_{\hat r})^\mu &=& \sqrt{\frac{\Delta}{\Sigma}}(0,1,0,0), \nn\\
(\e_{\hat \theta})^\mu &=& \sqrt{\frac{1}{\Sigma}} (0,0,1,0),\nn\\
(\e_{\hat \phi})^\mu &=& \sqrt{\frac{\Sigma}{\beta}}\frac{1}{\st}(0,0,0,1).
\een
and corresponding 1-forms are written as
\ben
\label{eq:1form}
(\e^{\hat t})_\mu &=& \sqrt{\frac{\Sigma\Delta}{\beta}}(1, 0, 0, 0), \nn\\
(\e^{\hat r})_\mu &=& \sqrt{\frac{\Sigma}{\Delta}}(0,1,0,0), \nn\\
(\e^{\hat \theta})_\mu &=& \sqrt{\Sigma} (0,0,1,0),\nn\\
(\e^{\hat \phi})_\mu &=& \sqrt{\frac{\beta}{\Sigma}}\st (-\frac{2ar}{\beta},0,0,1).
\een
Projecting the Faraday tensor $F_{\mu\nu}$ and its dual tenseor $*F_{\mu\nu}$ onto the tetrad in the way $\hat E_i = - F_{\mu\nu} (\e_{\hat t})^\mu  (\e_{\hat i})^\nu$ and $\hat B_i = *F_{\mu\nu} (\e_{\hat t})^\mu  (\e_{\hat i})^\nu$, we obtain the electric and magnetic fields measured by ZAMOs as \citep{Contopoulos2013, Nathanail2014, Penna2014}
\ben\label{eq:eb}
{\hat {\bm E}} &=& \frac{1}{\alpha\sqrt{\Sigma}}\left(\frac{2ar}{\beta}-\Omega\right)\left(\sqrt{\Delta}\Ar\ , -\st\Am\ , 0 \right), \nn\\
{\hat {\bm B}} &=& \frac{1}{\sqrt{\beta}\st}\left( -\st\Am\ , -\sqrt{\Delta}\Ar\ , -\frac{I\sqrt{\Sigma}}{\alpha}\right),
\een
where $\alpha=(\Delta\Sigma/\beta)^{1/2}$. 

An electromagnetic field that is generated by a spatially finite distribution of time-varying electric currents satisfies the radiation condition that electric field and magnetic field are of equal magnitude $\hat E = \hat B$ at far distance from the electric currents. While BH electrodynamics is special because its electric current $I$ is neither spatially confined nor time-varying, and the Poynting flux is nonzero only along magnetic field lines threading the event horizon. But the radiation condition still holds in the sense that $\hat E_\theta = \pm \hat B_\phi$ at infinity. At inner infinity $r\rightarrow r_+$, the Poynting flux measured by ZAMOs flows into the horizon, so we choose the ingoing radiation condition, $\hat E_\theta = - \hat B_\phi$, thus we have  
\be
I = \frac{(2r\Omega-a)\sst \Am}{\Sigma} \Big|_{r=r_+}, \nn
\ee 
which is exactly the Znajek regularity condition (\ref{eq:znajek}). At outer infinity $r\rightarrow\infty$, the Poynting flux measured by ZAMOs flows outwards, so we choose the outgoing radiation condition, $\hat E_\theta = \hat B_\phi$, consequently,
\be
\label{eq:inftyp}
I =  -\Omega \sst \Am\Big|_{r\rightarrow\infty},
\ee 
which confirms the generality of Eq.(\ref{eq:IOmega}) again, $I\propto\Omega$.\\

The constraint relations (\ref{eq:c1}-\ref{eq:c3}) can be reproduced by applying the above constraint to corresponding magnetic fields. For the asymptotically monopole fields, $A_\phi|_{r\rightarrow\infty}=-\mu$, Eq.(\ref{eq:inftyp}) is clearly equivalent to Eq.(\ref{eq:c1}). For the asymptotically uniform fields, $A_\phi|_{r\rightarrow\infty}=r^2\sst$, Eq.(\ref{eq:inftyp}) reads as
\be
I = 2\Omega A_\phi \mu\Big|_{r\rightarrow\infty},\nn
\ee
which is equivalent to Eq.(\ref{eq:c2}) though they seem to be different.
It is known that some magnetic field lines attach to the event horizon, and others do not. For the latter ones, we expect $I=\Omega=0$. For the former ones, the magnetic flux is finite $A_\phi\sim O(1)$, so $\theta\rightarrow0$ and $\mu\rightarrow 1$ in the limit $r\rightarrow\infty$. Hence we reproduce Eq.(\ref{eq:c2}), $I = 2\Omega A_\phi$, from the radiation condition Eq.(\ref{eq:inftyp}). Using the same argument, we can also reproduce Eq.(\ref{eq:c3}) from Eq.(\ref{eq:inftyp}) for the paraboloidal magnetic fields.

\section{Near-BH field lines}
\label{sec:field lines}

\begin{figure*}
\includegraphics[scale=0.54]{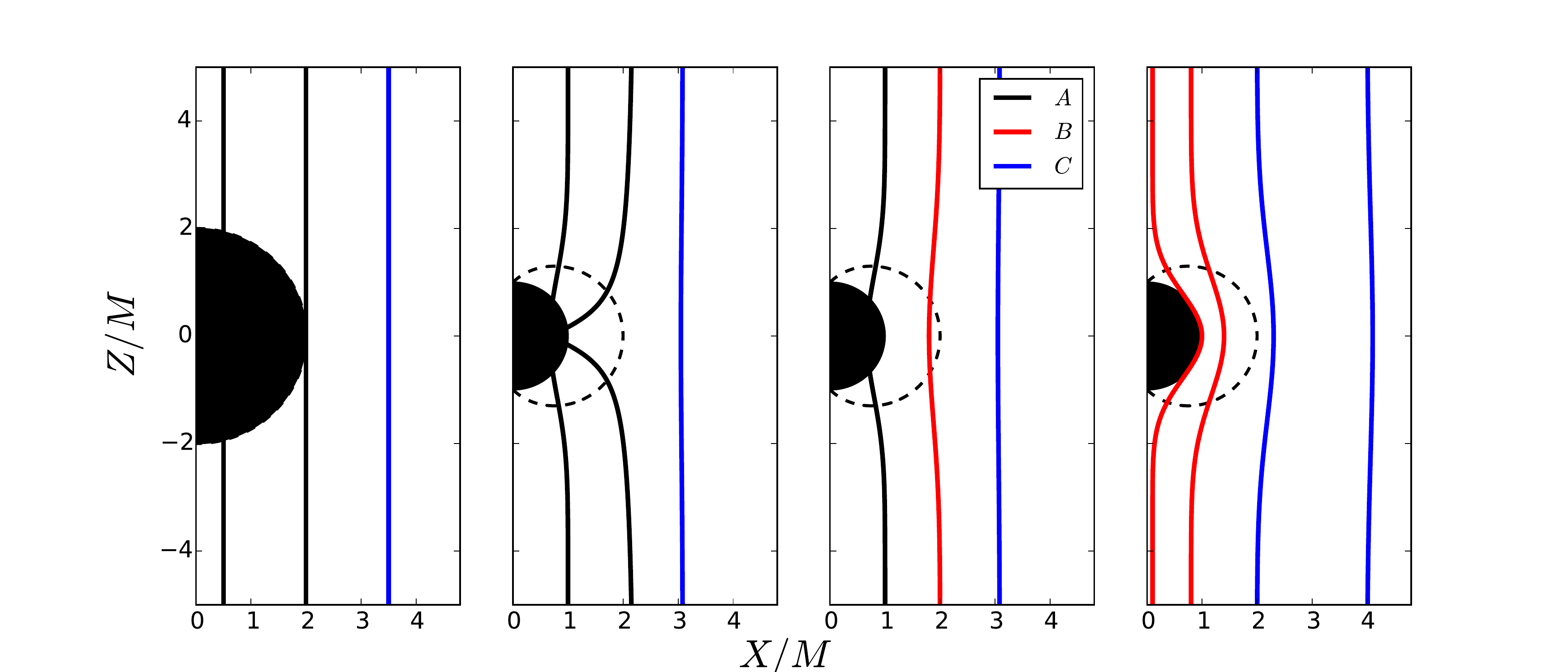}
\caption{Illustration of poloidal magnetic field lines in a BH magnetosphere, where the shadowed region in each panel is enclosed by the event horizon, and dashed curve is the infinite redshift surface, where $M$ is the gravitational radius of the central BH . \emph{First panel}: uniform vertical magnetic field around a Schwarzschild black hole. \emph{Second panel}: the configuration of magnetic field lines in a force-free (highly conductive) magnetosphere around an extremal Kerr BH. \emph{Third panel}: the configuration of magnetic field lines in a mildly conductive magnetosphere around an extremal Kerr BH.  \emph{Fourth panel}: the magnetic field configuration of the Wald vacuum (non-conductive) solution around an extremal Kerr BH.} 
\label{fig:x}
\end{figure*}

Based on the boundary conditions we have well understood, in this section, we will prove the following claim:

\begin{claim}
\label{claim}
In the steady axisymmetric force-free magnetosphere around a Kerr BH, all magnetic field lines that cross the infinite-redshift surface must intersect the event horizon.  
\end{claim}

Before proceeding to the proof, it necessary to emphasize that ``force-free" implies a perfectly conductive magnetosphere, which is completely different from a non-conductive magnetosphere, e.g., the magnetosphere of the vacuum solution constructed by \citet{Wald1974}. 

\begin{proof}

Without loss of generality, we take the asymptotically uniform vertical magnetic fields as an example \footnote{As demonstrated by \citet{Gralla2014}, ``\emph{A contractible force-free region of closed poloidal field lines cannot exist in a stationary, axisymmetric, force-free Kerr black hole magnetosphere}". Therefore all allowed configurations of magnetic field are topologically same to our example field. }.
In the \emph{first panel} of Figure \ref{fig:x}, we plot the poloidal field lines of uniform vertical fields in the magnetosphere of a non-rotating BH, and in the \emph{second/third/fourth panel} we plot the possible configurations of field lines distorted by the BH rotation. As shown in the plot, all field lines fall into three categories: type $\mathcal A$ (threading both the infinite-redshift surface and the event horizon), type $\mathcal B$ (threading the infinite-redshift surface only) and type $\mathcal C$ (threading none of them). 

For a non-rotating BH, the event horizon and the infinite-redshift surface coincide, so the proof is trivial. 
To prove Claim \ref{claim} for rotating BHs, we only need to prove that magnetic field lines of type $\mathcal B$ are forbidden by physical laws, which can be done by contradiction in two steps. \emph{Step 1}: If there was a  field line of type $\mathcal B$, the field line must cross the equator. \emph{Step 2}: the field line of type $\mathcal B$ cannot be force-free and cross equator at the same time.

\emph{Proof of step 1} :  It is known that a magnetic field line must close itself or ends at infinity. \emph {If the $\mathcal B$ type field line does not cross the equator} , there is no way for the field line to close itself, so it must terminate at infinity. One side of the field line ends at the outer infinity, and obviously the other side must end at the event horizon (the inner infinity). Thus it turns out be a type $\mathcal A$ field line (see \emph{second panel} of Figure \ref{fig:x}).  

\emph{Proof of step 2} : If the $\mathcal B$ type field line crosses  the equator (see \emph{third panel} of Figure \ref{fig:x}), according to the symmetry across the equator, the electric current either flows towards the equator from both $+z$ side and $-z$ side of the field line, or flows away from the equator to  infinity in both $+z$ direction and $-z$ direction. For each case, the electric charge conservation is violated. Therefore the only viable solution is vanishing poloidal electric current along the field line, i.e, $I=0$. According to the constraint condition [Eq.($\ref{eq:c2}$)], $I=2\Omega A_\phi$, the angular velocity would also vanish, $\Omega=0$. Now we need to consider plasmas attached to the field line. Due to the condition of perfect conductivity, plasmas attached to the field line corotate with it ($d\phi/dt = 0$).  While all non-rotating trajectories ($d\phi = 0$) within the ergosphere are spacelike $ds^2 > 0$ (easy to see from Eq.(\ref{eq:boyer}), also known as the frame-dragging effect), so the field line of type $\mathcal B$ is forbidden.   

To summarize, for a  type $\mathcal B$ field line, the symmetry requires vanishing poloidal electric current $I=0$, the frame-dragging effect requires $\Omega\neq0$, and the convergence constraint (or equivalent the radiation condition) requires $I\propto\Omega$. These three requirements cannot hold in the same time,  therefore  no type $\mathcal B$ line is allowed in a steady axisymmetric force-free BH magnetosphere. 
\end{proof}

The proof above is applicable to any steady axisymmetric force-free magnetosphere of other field configurations. More generally, simulations of both perfectly conductive, mildly conductive and non-conductive  BH magnetosphere were performed \citep{Komissarov2005,Komissarov2007d}.  It was found that in both force-free magnetosphere and highly conductive MHD magnetosphere, all magnetic field lines threading the infinite-redshift surface also thread the event horizon (\emph{second panel} of Figure \ref{fig:x}). For mildly conductive magnetosphere, some of the magnetic field lines escape from the event horizon into the ergosphere and some are trapped by the event horizon (\emph{third panel} of Figure \ref{fig:x}). While in non-conductive magnetosphere, rotating BHs tends to expel magnetic field lines out of the horizon (\emph{fourth panel} of Figure \ref{fig:x}). Such magnetic field expelling phenomena is similar to the Meissner effect in superconductors, so is called as the BH Meissner effect.

\section{BH Meissner effect}
\label{sec:meissner}
In this section, we first give a brief introduction to the Wald  solution \citep{Wald1974} as an example of the BH Meissner effect. Then we show that the geometry effect proposed by \citet{Penna2014} does not explain the BH Meissner effect. Finally, we argue that the Meissner effect is evaded by conductivity in a force-free BH magnetosphere.

\subsection{Wald solution}
The electromagnetic field of the Wald solution measured by ZAMOs is given by \citep{Frolov1998}
\be
\label{eq:wald}
{\hat {\bm B}} = \frac{1}{2\sqrt{\beta}\st}\left(X_{,\theta}, -\sqrt{\Delta} X_{,r}, 0\right),
\ee

\ben
{\hat {\bm E}}
= -\frac{a}{\Sigma}\sqrt{\frac{\beta}{\Delta}}\left\{\sqrt{\Delta}\left[ \alpha^2_{,r}
+X\left(\frac{r}{\beta}\right)_{,r}\right],\   \alpha^2_{,\theta}+X\left(\frac{r}{\beta}\right)_{,\theta},\  0 \right\} , \nn
\een
where $X=(\beta-4a^2r)\sst/\Sigma$, and we have chosen the magnetic field strength at infinity to be unity. On the constant $t$ hypersurface, total magnetic flux threading upper semi-sphere of the event horizon is expressed as
\ben
\Phi_{\rm H} 
&=& \int_{\partial C} \hat B_r \e^{\hat \theta}\wedge \e^{\hat\phi} = \int_{\partial C} \hat B_r \sqrt{\beta} \st d\theta d\phi \nn\\
&=& 4\pi \left(1-\frac{a^2}{r_+}\right),
\een
where we have used Eq.(\ref{eq:1form}, \ref{eq:wald}) and $dt=0$. So $\Phi_{\rm H}(a=1)=0$, i.e., all magnetic flux are expelled out of horizon for an extremal spinning BH (see the \emph{fourth panel} of Figure \ref{fig:x}, also previous works, e.g., \cite{King1975, Penna2014}). 

\subsection{Is the BH Meissner effect a geometric effect?}

\citet{Penna2014, Penna2014b} argued that the BH Meissner effect was a geometric effect that fields are unable to reach the horizon because the the length of the BH throat blows up in the extremal limit. Consider the magnetic flux threading the annulus right outside the event horizon at polar angle $\theta$, $\phi=(0, 2\pi)$, $r=(r_+, r_++\delta)$, 
\be
\Phi_\delta = \int \hat B_\theta \ \e^{\hat r}\wedge \e^{\hat\phi} = 2\pi \int_{r_+}^{r_++\delta} \hat B_\theta \sqrt{\frac{\beta}{\Delta}}\st dr,
\ee
where $\delta$ is an arbitrary displacement.
At the extremal limit, the right-hand side is finite only if $\hat B_\theta$ goes to zero at the horizon faster than or equal to $\sqrt{\Delta(r)}$. Hence \citet{Penna2014} concluded that the geometry effect ``rules out the possibility of steady axisymmetric fields threading the horizon in the extremal limit unless the field becomes entirely radial at the horizon''. 

The above conclusion is correct but irrelevant, because for any field configurations, the $\theta$ component of the magnetic field approaches to zero in the way $\hat B_\theta \sim \sqrt{\Delta(r)}$ [Eq.(\ref{eq:eb}, \ref{eq:wald})] in the vicinity of the event horizon, and consequently poloidal field lines lie in the direction of $\e^{\hat r}$ on the horizon. Therefore no magnetic field configuration is ruled out by the finite $\Phi_\delta$ requirement or equivalently the requirement of vanishing $\hat B_\theta$ at the horizon.  More specifically, for force-free solutions [Eq.(\ref{eq:eb})], the magnetic flux threading the upper semi-sphere of the event horizon is
\be
\Phi_{\rm H} = 2\pi A_\phi\Big|_{\theta=0}^{\theta=\pi/2},
\ee
and the flux threading the annulus is 
\be
\Phi_\delta = 2\pi A_\phi\Big|_{r_++\delta}^{r_+}.
\ee
It is straightforward that $\Phi_\delta$ is finite for any finite $A_\phi$. Therefore the finite $\Phi_\delta$ requirement imposes no constraint on $A_\phi$ or the magnetic flux threading the horizon $\Phi_{\rm H}$, hence is irrelevant to the BH Meissner effect.\\

\subsection{The BH Meissner effect does not work in a force-free magnetosphere}
The BH Meissner effect emerges in the Wald vacuum solution and other steady axisymmetric vacuum solutions. It is especially important to examine whether it operates in real astrophysical environment where BH magnetosphere are highly conductive and force-free. 

Assuming the BH Meissener effect also works in the force-free magnetosphere: all magnetic field lines are expelled out of the horizon of an extremal Kerr BH. Combining Claim \ref{claim} we proved in Section \ref{sec:field lines}, it is known that field lines penetrating the infinite redshift surface must also penetrate the event horizon, so all field lines would be expelled out of the ergosphere, i.e.,  only field lines of type $\mathcal C$ exist (see Figure \ref{fig:x}). But such field configuration is neither natural nor stable. There is no any mechanism that would forbid electromagnetic fields within the ergosphere, so the configuration is not natural. Even a magnetic field line initially located outside the ergosphere, it would be attracted into the ergosphere ( see the simulations of a single magnetic flux tube in the Kerr spacetime done by \citet{Semenov2004} ). As long as a  field line enters the ergosphere,  it must terminate at the event horizon. Therefore magnetic field lines in a force-free magnetosphere will not be expelled by BH rotation. 

People may wonder the difference between the force-free solution (see the \emph{second panel} of Figure \ref{fig:x}) and the Wald vacuum solution (see the \emph{fourth panel} of Figure \ref{fig:x}).  For the force-free solution, the poloidal electric current $I$ is generated by the frame-dragging effect and the perfect conductivity, which splits all field lines of type $\mathcal B$ at the equator; the split field lines must terminate at the event horizon \footnote{According to \citet{Komissarov2007d}, the phenomena that magnetic field lines are attracted to instead of expelled out of the event horizon in a force-free magnethosphere is a result of hoop stress. The poloidal electric current $I$ produces toroidal magnetic fields whose hoop stress tends to pull the magnetic flux back to the event horizon.}. For the Wald vacuum solution, poloidal electric current $I$ is zero, therefore field lines of type $\mathcal B$ are allowed to exist, and so it is possible for field lines to bypass the horizon.  In a word, the essential difference between the two is poloidal electric current \citep{Gralla2014} or conductivity \citep{Komissarov2005,Komissarov2007d}.

\section{Conclusion}
\label{sec:conclusion}
The structure of a  steady axisymmetric force-free BH magnetosphere is governed by the GS equation (\ref{eq:GS}), which is a second-order differential equation of $A_\phi(r,\theta)$ and depends on two free functions $I(A_\phi)$ and $\Omega(A_\phi)$. To determine a solution $A_\phi(r,\theta)$, we need to specify $I(A_\phi)$ and $\Omega(A_\phi)$ self-consistently. In this paper, we apply the perturbation approach proposed by \citet{Pan2015a} to various magnetic field configurations, and sort out two self-consistent boundary conditions for each field configuration: Znajek regularity condition at inner infinity (the event horizon)  and convergence requirement at outer infinity, where the outer boundary condition could be casted to be a simple constraint condition in the form of $I=\Omega\times \mathcal{F}(A_\phi)$ [see Eqs. (\ref{eq:c1}-\ref{eq:c3})]. We interpret the two conditions mathematically and physically. Mathematically the two conditions are results of convergence requirement and physically they are  radiation conditions at inner and outer infinity. 

Based on the above two boundary conditions, we prove the claim that in a steady axisymmetric  force-free BH magnetosphere, all magnetic field lines threading the infinite-redshift surface must terminate at the event horizon. It  naturally explains that jets in various GRMHD simulations spontaneously develop a large split-monopole component in the vicinity of the central spinning BHs despite the initial field configurations  (see the second panel of Figure \ref{fig:x}).  We further argue that the BH Meissner effect does not operate in a force-free BH magnetosphere due to the perfect conductivity. Therefore the BZ mechanism would not be undermined in real astrophysical environment where BH magnetosphere are highly conductive.\\

C.Y. is grateful for the support by the National Natural Science Foundation
of China (Grant 11173057, 11373064, 11521303), Yunnan Natural Science
Foundation (Grant 2012FB187, 2014HB048). Part of the computation was performed
at the HPC Center, Yunnan Observatories, CAS, China.
This work made extensive use of the NASA Astrophysics Data System and
of the {\tt astro-ph} preprint archive at {\tt arXiv.org}.


\end{document}